\documentclass[1p]{elsarticle}
\usepackage{epsfig}
\usepackage{color}
\usepackage{epsfig}

\begin{document}

\begin{frontmatter}
\title{Majority vote model with ancillary noise in complex networks}

\author[if-USP]{J. M. Encinas}
\author[Anhui]{Hanshuang Chen}
\author[UFSJ]{Marcelo M. de Oliveira}
\author[if-USP]{Carlos E. Fiore}
\ead{fiore@if.usp.br}


\address[if-USP]{Instituto de F\'{\i}sica,
Universidade de S\~{a}o Paulo, \\
Caixa Postal 66318, 05315-970 S\~{a}o Paulo, S\~{a}o Paulo, Brazil}

\address[Anhui]{School of Physics and Material Science, Anhui University, Hefei 230039, China}

\address[UFSJ]{Departamento de F\'{\i}sica e Matem\'atica,
CAP, Universidade Federal de S\~ao Jo\~ao del Rei,\\
36420-000 Ouro Branco-MG, Brazil}

\date{\today}

\begin{abstract}
   We analyze the properties of the majority-vote (MV) model with
  an additional noise in which a local spin can be changed
 independently of its neighborhood. In the standard MV, one of the simplest nonequilibrium
  systems exhibiting  an order-disorder phase transition,
spins  are  aligned with their local majority 
with probability $1-f$, and with complementary probability $f$, 
the majority rule is not followed. In the noisy MV (NMV), a 
random spin flip is succeeded with probability $p$ (with  complementary
$1-p$ the usual MV rule is accomplished). Such extra ingredient
  was considered by Vieira and Crokidakis [Physica A {\bf 450},
  30 (2016)]  for the square lattice. 
 Here, we generalize the NMV for arbitrary networks, including
homogeneous [random regular (RR) and Erd\"os Renyi (ER)] and heterogeneous
[Barabasi-Albert (BA)] structures, through mean-field calculations and numerical simulations.
Results coming from both approaches are in excellent
agreement with each other, revealing  
that the presence of additional noise does not affect
  the classification of phase transition, which remains 
 continuous irrespective of the network degree and its distribution. 
 The critical point and  the threshold
  probability $p_t$ marking the disappearance of  the ordered phase 
depend on the node distribution and increase with the connectivity $k$.
The critical behavior, investigated numerically,  exhibits a common set
of critical exponents for RR and ER topologies, but different from BA
and  regular lattices.
 Finally, our results indicate that (in contrary to a previous proposition) there is no first-order transition in the NMV
for large $k$.

\end{abstract}

\end{frontmatter}

\section{Introduction}
 Phase transitions and spontaneous breaking
symmetry appear in a myriad of systems in the scope
of physics \cite{marr99,odor07,mario92,qvoter}, biology \cite{vicsek}, chemistry, social dynamics
 \cite{social,social2} and others. 
  Among the several microscopic models that  exist in the literature, the Majority Vote (MV) model is probably one of the simplest nonequilibrium models exhibiting up-down $Z_2$ symmetry \cite{mario92}.  
 Its dynamics  mimic the existence/formation of
 different opinions ($\pm 1$ in such case)
 in a community. In particular, it includes the role of individuals that
 do not adopt the local prevailing (majority)  opinion,
 commonly referred to as contrarians  or nonconformist characters.
  Therefore, it has attracted
considerable attention recently, not only for theoretical purposes (including
the investigation of critical behavior, universality classes, and phase coexistence) 
 but also for the description (at least
  in a reduced level) of the mechanisms leading to the opinion
 formation \cite{social,social2}.
It is known that the phase transition exhibited by the MV model
is continuous, signed by a spontaneous symmetry-breaking  
\cite{mario92,chen1,pereira}, although
the critical behavior depends on lattice topology  \cite{mario92,pereira}.

An entirely different behavior
was recently uncovered \cite{chen2,encinas} with the
inclusion of  a term proportional to the local spin (an inertial term), in which
the phase transition becomes discontinuous. 
The effects of other ingredients, such as {\em partial} inertia \cite{harunari},
more states per site (instead of ``up'' or ``down'' 
as in the MV) \cite{chen2,three-state}, and
diffusion \cite{diffusion} have also been investigated.

The effect of an {\em ancillary} noise in the MV model 
 was first studied  by Vieira and Crokidakis \cite{crokidakis}. 
In such noisy MV model (NMV), besides the intrinsic noise $f$,
ruling the majority interaction  to be or not to be accepted, 
one includes an independent term
allowing a spin  to be flipped irrespective of its neighborhood. In the social
language jargon, this ingredient corresponds to another kind of nonconformism,
usually referred to as {\em independence}.
 It was found a critical behavior  in the square lattice
  identical to the that observed in the standard MV model, although
the presence of extra noise provides an additional route for
the emergence of the phase
transition, even in the absence of misalignment term $f$ \cite{crokidakis}.

A relevant issue concerns in comparing the NMV 
with  similar models in which the effect of independent noise has been undertaken. 
Recent studies for the $q$-voter model  \cite{qvoter}, in which  a spin
is flipped to the value of its $q$ nearest neighbors with a certain
rate, reveal that the transition becomes discontinuous for large values of $q$,
provided the independence ingredient is included. 
Due to the similarities between the $q-$voter model and the NMV,  
a natural question that arises is if that shift in
the order of the phase transition is also verified in the NMV
for high connectivity \cite{crokidakis}.

With these ideas in mind, here we  analyze the NMV
 in distinct topologies for arbitrary connectivity.
 Through of  mean-field calculations (MFT), we 
derive expressions for  the critical
points as a function of the extra noise 
for an  arbitrary network topology.
We employ a different (MFT) approach than Refs. \cite{chen1,romualdo},
which is based on an
extension of the ideas from Ref. \cite{mario92} for an arbitrary network structure. The critical behavior is also investigated by performing
numerical simulations. 
 Although MFT  provides an approximate
description of the phase transition, we observe an 
excellent agreement with numerical results in the regime of high 
connectivities. Our findings reveal the phase transition
is continuous for all topologies and connectivities studied.
Thus,  contrasting to the $q-$voter model, a first-order 
transition is not observed in the NMV.  Finally, our upshots 
also indicate the critical
behavior in scale-free networks is different from that observed in 
homogeneous networks.

This paper is organized as follows:  In Sec. II, we introduce the model and perform
a mean field treatment. In Sec.III, we show and discuss our numerical results. 
At last, conclusions are drawn in Sec. IV.

\section{Model and mean field analysis}

The  MV model is defined in 
an arbitrary lattice topology, in which
each site $i$ of degree $k$ is attached to a
binary spin variable, $\sigma_i$,  that can take the values 
$\sigma_i=\pm 1$.  In the original model, with probability $1-f$
each node $i$ tends to align itself with its local neighborhood majority,
 and with complementary probability $f$, the majority rule is not
followed.  The increase of the misalignment quantity $f$ gives rise to an order-disorder (continuous) phase transition \cite{mario92,pereira,chen1}.
The NMV differs from the original MV due to the inclusion of  spin inversion
independently of the neighborhood.
Mathematically, it  is equal to the following  transition rate
\begin{equation}
         w(\sigma_i)=\frac{(1-p)}{2}[1-(1-2f)\sigma_{i}
         S(\sum_{j=1}^{k}\sigma_j)]+\frac{p}{2},
\label{eq2}
\end{equation}
where $S(X)$ is defined by $S(X)={\rm sign (X)}$  
if $X \neq 0$ and $S(0)=0$. 
Note that for $p=0$ one recovers
the original MV model.
Since we are dealing with Markovian systems ruled by a master equation,
our analysis starts by deriving  
the time evolution of the local magnetization of a site $i$ with degree $k$, $m_k=\langle \sigma_i\rangle_k$,   given by 
$d \langle \sigma_i\rangle_k/dt=-\langle 2\sigma_{i}w(\sigma_i)\rangle_k$. From the transition rate in Eq. (\ref{eq2}), we have that
\begin{equation}
  \frac{1}{1-p}\frac{d }{dt}m_k=-m_k+(1-2f)\langle S(X) \rangle-\frac{p}{1-p}m_k.
\label{eq3}
\end{equation}

Our first inspection regarding the extra noise effect is carried out
through a mean-field treatment. We shall consider a somewhat
different approach (although equivalent) than Refs. \cite{chen1,romualdo},
in which  the  mean sign function is decomposed in two parts,
$\langle S(X)\rangle=\langle S(X_+)\rangle-\langle S(X_-)\rangle$, with
each term $\langle S(X_i)\rangle$ being approximated as
\begin{equation}
  \langle S\left(X_\pm\right)\rangle \approx \pm\sum_{n=\lceil k/2 \rceil}^{k}C_{n}^{k}p_{\pm }^{n}p_{\mp }^{k-n}.
  \label{eq4}
\end{equation}
  Here,  each term $C_{n}^{k}$ in $\langle S\left(X_\pm\right)\rangle$ takes into account the number of
possibilities of a neighborhood with $n$ and $k-n$ spins
in the $+1(-1)$ and $-1(+1)$ states 
with associated probabilities $p_{+}(p_-)$ and $p_-(p_+)$, respectively. 
In the following, we shall relate $p_{\pm }$ to the 
 local magnetization $m_k$. 
  By focusing our analysis on uncorrelated structures, the probability $\rho_k$ of a local site 
presenting the spin $+1$ is related to
$p_{\pm }$ through 
$p_{+}=\sum_{k}kP(k)\rho_k/\langle k \rangle $,
where $P(k)$ is the degree probability distribution
with mean value $\langle k \rangle$.
Recalling that $\rho_k$ and $m_k$ are constrained through $m_k=2\rho_k-1$,
the steady solution of Eq. (\ref{eq3})   can be rewritten
solely in terms of  $p_{\pm}$ given by
\begin{equation}
  p_+-\frac{1}{2}=\frac{(1-2f)}{2\langle k\rangle}\sum_{k=1}^{\infty}kP(k)\sum_{n=\lceil k/2 \rceil}^{k}C_{n}^{k}(p_{+}^{n}p_{-}^{k-n}-p_{-}^{n}p_{+}^{k-n})-\frac{p}{1-p}(p_+-\frac{1}{2}).
   \label{eq4-1}
\end{equation}
It is possible to derive a simpler expression for
$\langle S\left(X_\pm\right)\rangle$ for large  $k$s, in which each term from the binomial distribution  approaches to a Gaussian one
with mean $kp_{\pm }$ and  variance $\sigma^{2}=kp_{+}p_{-}$.
So that  $\sum_{n=\lceil k/2 \rceil}^{k}C_{n}^{k}p_{\pm}^{n}p_{\mp}^{k-n}$
becomes
$\int_{k/2}^k d\ell e^{-(\frac{\ell-kp_{\pm}}{\sqrt{2}\sigma})^2}/(\sigma\sqrt{2\pi})=\frac{\sqrt{\pi}}{2}({\rm erf}[\frac{k(1-p_{\pm})}{\sqrt{2}\sigma}]-{\rm erf}[\frac{k(1/2-p_{\pm})}{\sqrt{2}\sigma}])$, where ${\rm erf(x)}=2\int_{0}^{x}e^{-t^2}dt/\sqrt{\pi}$ denotes the error function.
Thereby, the expression for the mean $\langle S(X)\rangle$ reads 
\begin{equation}
\langle S(X)\rangle=\frac{1}{2}\left[2{\rm erf}\Bigg(\frac{\sqrt{2k}y}{\sqrt{1-4y^2}} \Bigg) + {\rm erf}\Bigg(\frac{\sqrt{2k}(\frac{1}{2}-y)}{\sqrt{1-4y^2}} \Bigg) - {\rm erf}\Bigg(\frac{\sqrt{2k}(\frac{1}{2}+y)}{\sqrt{1-4y^2}} \Bigg)  \right],
\label{eq5}
\end{equation}
 where $p_{\pm}$ is  rewritten in terms of the variable 
 $y$  through the formula $p_{\pm}=\frac{1}{2}\pm y$ and for large
 $k$ Eq. (\ref{eq4-1}) becomes
 \[
   y=\frac{(1-2f)}{4\langle k\rangle}\sum_{k=1}^{\infty}kP(k)\left[2{\rm erf}\Bigg(\frac{\sqrt{2k}y}{\sqrt{1-4y^2}} \Bigg) + {\rm erf}\Bigg(\frac{\sqrt{2k}(\frac{1}{2}-y)}{\sqrt{1-4y^2}} \Bigg)-{\rm erf}\Bigg(\frac{\sqrt{2k}(\frac{1}{2}+y)}{\sqrt{1-4y^2}} \Bigg)\right]+  \]
\begin{equation}
  -\frac{p}{1-p}y.
 \label{eq4-2}
\end{equation}

  Note that for large $k$ the numerator dominate over the denominator, so
the second and third  terms in the right side of
Eqs. (\ref{eq5}) and (\ref{eq4-2}) cancel themselves and
 $\langle S(X)\rangle$ reduces to the simpler form
 $\langle S(X)\rangle={\rm erf}(y\sqrt{2k})$, from which we arrive at the
  following  steady-state relation:
\begin{equation}
y=\frac{1}{2\langle k \rangle}(1-2f)\sum_{k}[k{\rm erf} (y\sqrt{2k})]P(k)-\frac{p}{1-p}y.
\label{eq7}
\end{equation}
Thus, having the steady $y_0$s [from Eqs. (\ref{eq4-1}) and (\ref{eq7})],
the correspondent $m_k$ are then given by (for fixed $f$ and $p$)
\begin{equation}
m_k=(1-2f){\rm erf}(y\sqrt{2k})-\frac{p}{1-p}m_k,
\label{eq6-1}
\end{equation}
whose mean
magnetization $m$ is finally evaluated through
$m=\sum_{k=1}^{\infty}m_kP(k)$.

In order to derive a closed expression
for the critical point, we 
should note that Eq. (\ref{eq7}) presents two solutions 
($y=\pm y_0 \neq 0$) for $f<f_c$ (besides the trivial $y=0$) 
and only the trivial solution $y=0$ for $f>f_c$. 
Since $y_0$ is expected to be small close to the critical point,
 the first term in the  right side of Eq. (\ref{eq7})  can be expanded in Taylor series
whose dependence on $f$ and $p$  (for arbitrary lattice topology) reads
\begin{equation}
  y_0=\left\{\frac{1}{A(f,p)}\left[-1+(1-p)(1-2f)\sqrt{\frac{2}{\pi}}\frac{\langle k^{3/2} \rangle}{\langle k \rangle} \right]\right\}^{1/2},
\label{eq7-1}
\end{equation}
where $\langle k^{3/2} \rangle=\sum_{k}k^{3/2}P(k)$ and 
$A(f,p)=\sqrt{\frac{8}{9\pi}}\frac{\langle k^{5/2}\rangle}{\langle k\rangle}(1-p)(1-2f)$.  Finally, the critical $f_c$ is given by 
\begin{equation}
  f_c=\frac{1}{2}-\frac{1}{2}{\sqrt\frac{\pi}{2}}\frac{1}{1-p}\frac{\langle k \rangle}{\langle k^{3/2} \rangle},
\label{eq8}
\end{equation}
with $A(p,f_c)=2\langle k^{5/2} \rangle/\langle k^{3/2} \rangle$.
In particular, for $p=0$ one recovers the expression $f_c=\frac{1}{2}-\frac{1}{2}{\sqrt\frac{\pi}{2}}\frac{\langle k \rangle}{\langle k^{3/2} \rangle}$,
in consistency with the results from Ref. \cite{chen1}.
Complementary, 
an order-disorder phase transition is also obtained
in the absence of misalignment $f=0$ by  increasing
the ancillary noise $p$, whose critical rate $p_c$
satisfies the relation
$p_c=1-({\sqrt\pi}\langle k \rangle)/({\sqrt 2}\langle k^{3/2} \rangle)$.
Thus, above expression extends  
the conjecture  $p_c(f=0)=2f_c(p=0)$,
obtained  for the square lattice \cite{crokidakis},
for arbitrary  network distribution.
A second MFT upshot to be drawn  is that,  contrasting to the 
generalized $q-$voter model \cite{qvoter}, 
 the inclusion  of  an independent noise
 does not alter the classification of phase transition,
 irrespective of the lattice topology and the neighborhood. The critical point,
on the other hand,  depends on the network topology.

 The first structure we consider is a random regular 
network in which  nodes follow the distribution
$P(k)=\delta (k-k_0)$, with $k_0$ being the degree.  In this case, Eq. (\ref{eq8}) becomes
\begin{equation}
  f_c=\frac{1}{2}-\frac{1}{2}{\sqrt\frac{\pi}{2k_0}}\frac{1}{1-p}.
  \label{rrr}
  \end{equation}
The second topology considered is the Erd\"os-Renyi (ER) network, an iconic example
 of a homogeneous random network, with
a degree distribution given by
$P(k)=\langle k \rangle^{k}e^{-\langle k \rangle}/k!$.
Finally, a heterogeneous network, in which
nodes are distributed according to a power-law distribution 
$P(k) \sim k^{-\gamma}$, is considered. For avoiding divergences
when $k \rightarrow 0$ in the PL, 
we have imposed  a minimum degree $k_0$  and the averages $\langle k\rangle$
and $\langle k^{3/2}\rangle$ become 
$\langle k\rangle=(\gamma-1)k_0/(\gamma-2)$
and
$\langle k^{3/2}\rangle=(\gamma-1)k_0^{3/2}/(\gamma-5/2)$,
respectively. Thus, the critical point $f_c$ reads
\begin{equation}
 f_c=\frac{1}{2}-\frac{1}{2}{\sqrt\frac{\pi}{2k_0}}\frac{\gamma-5/2}{\gamma-2}\frac{1}{1-p}. 
\label{ucmr}
\end{equation}
  In this work, we shall focus on the analysis for $\gamma=3$, which is a hallmark
of scale-free structures \cite{ba}.
In the next section,
we are going to confirm above findings by performing numerical simulations.

\section{Numerical Results and phase diagrams}

 We have performed extensive numerical simulations
for networks with sizes $N$ ranging from $N=1000$ to $20000$. 
We generated the RR networks through the classical
configuration model, 
introduced by Bollob\'as \cite{boll}.
The ER networks are constructed by connecting each pair of nodes with probability $\langle k \rangle/N$.
 When the size of the graph $N\rightarrow \infty$,   
the degree distribution is  Poissonian,
with mean $\langle k \rangle$. The Barabasi-Albert (BA) scale-free network
is a typical representation of heterogeneous structures \cite{ba}, in
which the degree distribution follows a power-law $P(k) \sim k^{-\gamma}$
with scaling exponent $\gamma=3$.

Besides the order parameter vanishment,
continuous phase transitions are signed by an algebraic divergence
 of the variance $\chi=N[\langle m^{2} \rangle-|m|^{2}]$
 at the critical point $f_c$ for $N\rightarrow \infty$ \cite{marr99,odor07,henkel}.  Since
 only finite systems can be simulated,  these quantities 
become rounded at the vicinity of criticality due to finite size effects.
 For calculating  the critical point and
the critical exponents, we resort to the finite size 
scaling theory,
 in which  $|m|$ and  $\chi$ are rewritten as $|m|=N^{-\beta/\nu}{\tilde f}(N^{1/\nu}|\epsilon|)$ and $\chi=N^{\gamma/\nu}{\tilde g}(N^{1/\nu}|\epsilon|)$, with
  ${\tilde f}$ and ${\tilde g}$ being scaling functions and 
$\epsilon=(f-f_c)/f_c$. For $\epsilon=0$, the above relations acquire
the dependence on the system size reading $|m|=N^{-\beta/\nu}{\tilde f}(0)$ and $\chi=N^{\gamma/\nu}{\tilde g}(0)$, in which a log-log plot
of $|m|$ and $\chi$ versus $N$ furnish the exponents 
$\beta/\nu$ and $\gamma/\nu$, respectively.
\begin{figure}[h!]
\centering
\epsfig{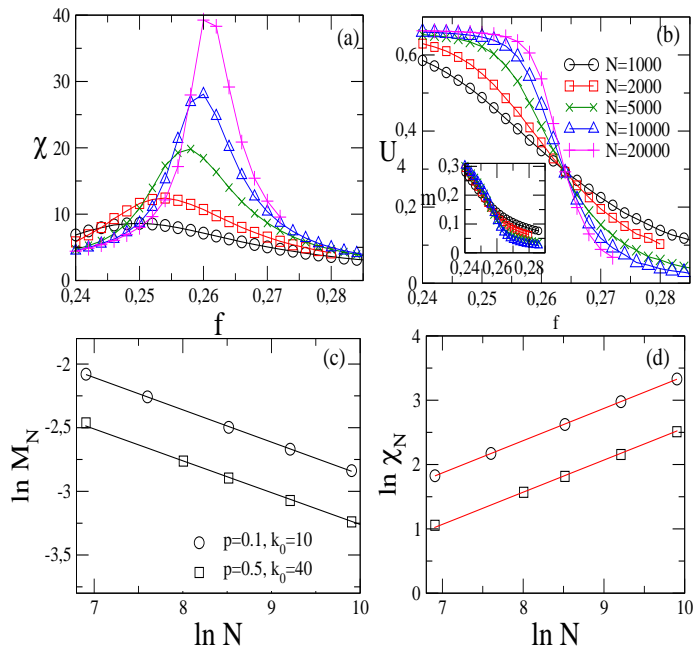}
\caption{Panels $(a)$ and $(b)$ show, for the RR networks and $p=0.1$ and $k_0=10$,
  the variance $\chi$ and the reduced cumulant $U_4$ versus
  $f$ for distinct system sizes $N$. Inset: the magnetization
  per spin $m$ vs. $f$. Panels $(c)$ and $(b)$ show the critical exponents
$\beta/\nu$ and $\gamma/\nu$ for $p=0.1$ ($k_0=10$) and $p=0.5$($k_0=40$), respectively. They are consistent with  $\beta/\nu=1/4$ and $\gamma/\nu=1/2$,
respectively.}
\label{fig1}
\end{figure}
The critical point can be properly located
through the reduced cumulant
$U_4=1-\langle m^{4} \rangle/(3\langle m^{2} \rangle^{2})$,
because curves for distinct  $N$s cross at   $f=f_c$ $(\epsilon=0)$ 
and $U_4$ becomes constant $U_4=U_0^{*}$. Off the critical
point, $U_4 \rightarrow 2/3$ and $0$
for the ordered and disordered phases, respectively when $N \rightarrow\infty$.

\begin{figure}[h!]
\centering
\epsfig{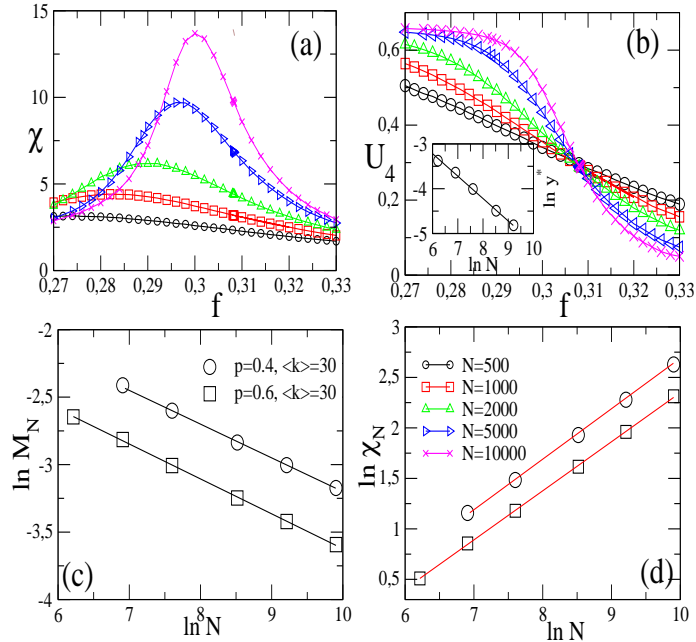}
\caption{Panels $(a)$ and $(b)$  show, for the ER case, $p=0.4$ and
  $\langle k\rangle=30$,
  the variance $\chi$ and the reduced cumulant $U_4$ versus
  $f$ for distinct system sizes $N$. Inset:
  Log-log plot of $y^*\equiv f_c-f_N$ versus $N$, with  $f_N$
  calculated from the maximum of $\chi$. The straight line has
  slope consistent to $1/\nu=1/2$. Panels $(c)$ and $(b)$ show the critical exponents
  $\beta/\nu$ and $\gamma/\nu$ for $p=0.4$  and
  $p=0.6$ (both for $\langle k\rangle=30$), respectively.  
They are consistent with  $\beta/\nu=1/4$ and $\gamma/\nu=1/2$, respectively.}
\label{fig1-2}
\end{figure}

Figs. \ref{fig1} and \ref{fig1-2} exemplify the behavior of  above
 quantities for the RR  and ER cases
for  $p=0.1$ ($k_0=10$), $p=0.5$ ($k_0=40$) 
and $p=0.4$, $0.6$ ($\langle k\rangle=30$), respectively.
Panels $(a)$ and $(b)$ reproduce  
the typical trademarks of critical transitions: 
$|m|$ decreases smoothly by raising $f$ (or $p$) 
and $\chi$ presents a  maximum 
whose peak becomes more pronounced  as  $N$ increases.
The nature of  the phase transitions
is reinforced by examining the crossing among the curves of $U_4$
for distinct system sizes. In all cases, the crossing
 is characterized by an apparent universal value $U_0^*=0.28(1)$.

Analysis of the critical exponents furnish results consistent 
with  $\beta/\nu=1/4$, $\gamma/\nu=1/2$ 
and $1/\nu=1/2$ (see inset of Fig. \ref{fig1-2}).   Also, they satisfy the
relation $2\beta/\nu+\gamma/\nu=D_{eff}$, with $D_{eff}=1$,
indicating a universal behavior for homogeneous topologies.
On the other hand, all of them are very different from 
the values $\beta=1/8$, $\gamma=7/4$ 
and $\nu=1$ for bidimensional  (regular) lattices \cite{crokidakis,mario92}.
\begin{figure}[h!]
\centering
\epsfig{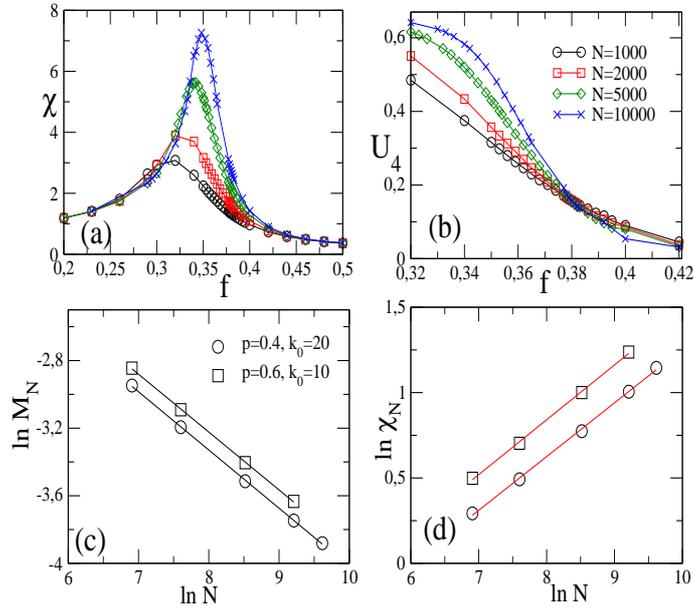}
\caption{Panels $(a)$ and $(b)$  show, for the BA network and $p=0.4$ and
  $k_0=20$,
  the variance $\chi$ and the reduced cumulant $U_4$ versus
  $f$ for distinct system sizes $N$.  Panels $(c)$ and $(b)$ show the critical exponents
  $\beta/\nu$ and $\gamma/\nu$ for $p=0.4$ ($k_0=20$) and
  $p=0.6$ ($k_0=10$), respectively.  
They are consistent with  $\beta/\nu=0.34(1)$ and $\gamma/\nu=0.32(1)$, respectively.}
\label{fig1-3}
\end{figure}

 A slightly distinct critical behavior is obtained for the BA network,
exemplified in Fig. \ref{fig1-3} for $p=0.4$ ($k_0=20$) and $p=0.6$ ($k_0=10$).
In both cases, the crossing value
$U_0^*$ and  the set of critical exponents are
other than those
obtained for the homogeneous structures, reading
$U_0^*=0.16(1)$ and critical exponents
$\beta/\nu=0.34(1)$ and $\gamma/\nu=0.32(1)$, respectively.
{Albeit, they also fulfill
the relation $2\beta/\nu+\gamma/\nu=1$.}
\begin{figure}[h!]
\centering
\epsfig{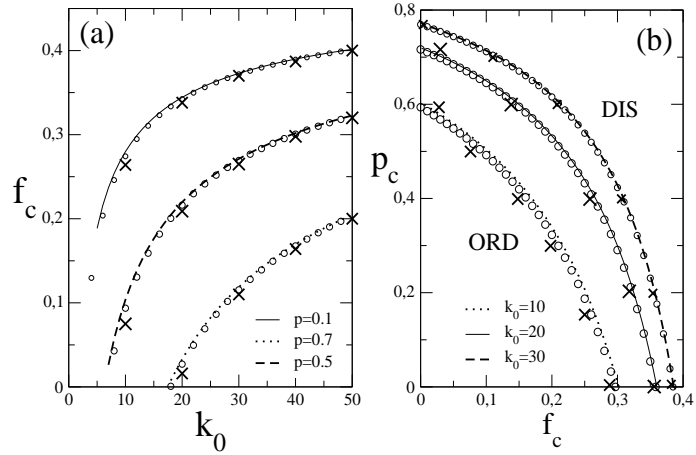}
\caption{Panel $(a)$ shows, for the RR network,
  the transition rates $f_c$ versus the degree node $k_0$
  for distinct values of $p$. Circles and  lines  correspond
  to the estimates obtained from Eqs. (\ref{eq4-1}) and (\ref{rrr}), respectively. The
  symbol $\times$ correspond to the numerical values obtained from  the crossing among $U_4$ for distinct $N$s. Panel  $(b)$  shows the phase diagram $p_c$ versus $f_c$ for distinct
$k_0$s.}
\label{fig2}
\end{figure}

\begin{figure}[h!]
\centering
\epsfig{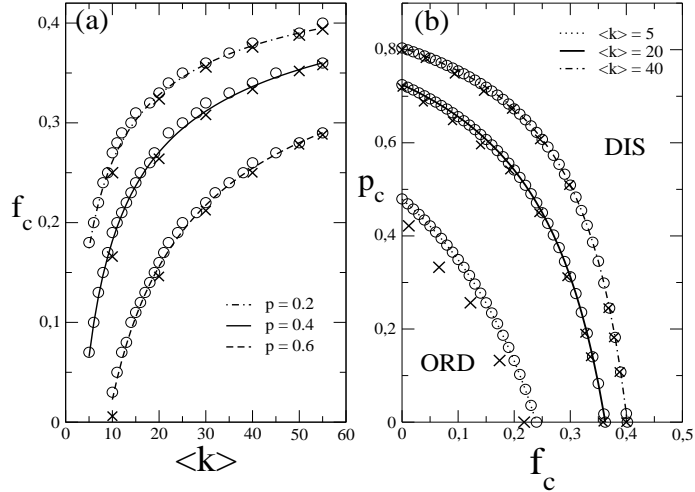}
\caption{Panel $(a)$ shows, for the ER case,
  the transition rates $f_c$ versus the mean degree $\langle k\rangle$
  for distinct values of $p$. Circles and  lines correspond
  to estimates obtained from Eqs. (\ref{eq4-1}) and (\ref{eq8}), 
  respectively.
The
  symbol $\times$ correspond to the numerical values obtained from  the crossing among $U_4$ for distinct $N$s. Panel $(b)$ shows the phase diagram $p_c$ versus $f_c$ for distinct
$\langle k\rangle$s.}
\label{fig3}
\end{figure}

\begin{figure}[h!]
\centering
\epsfig{file=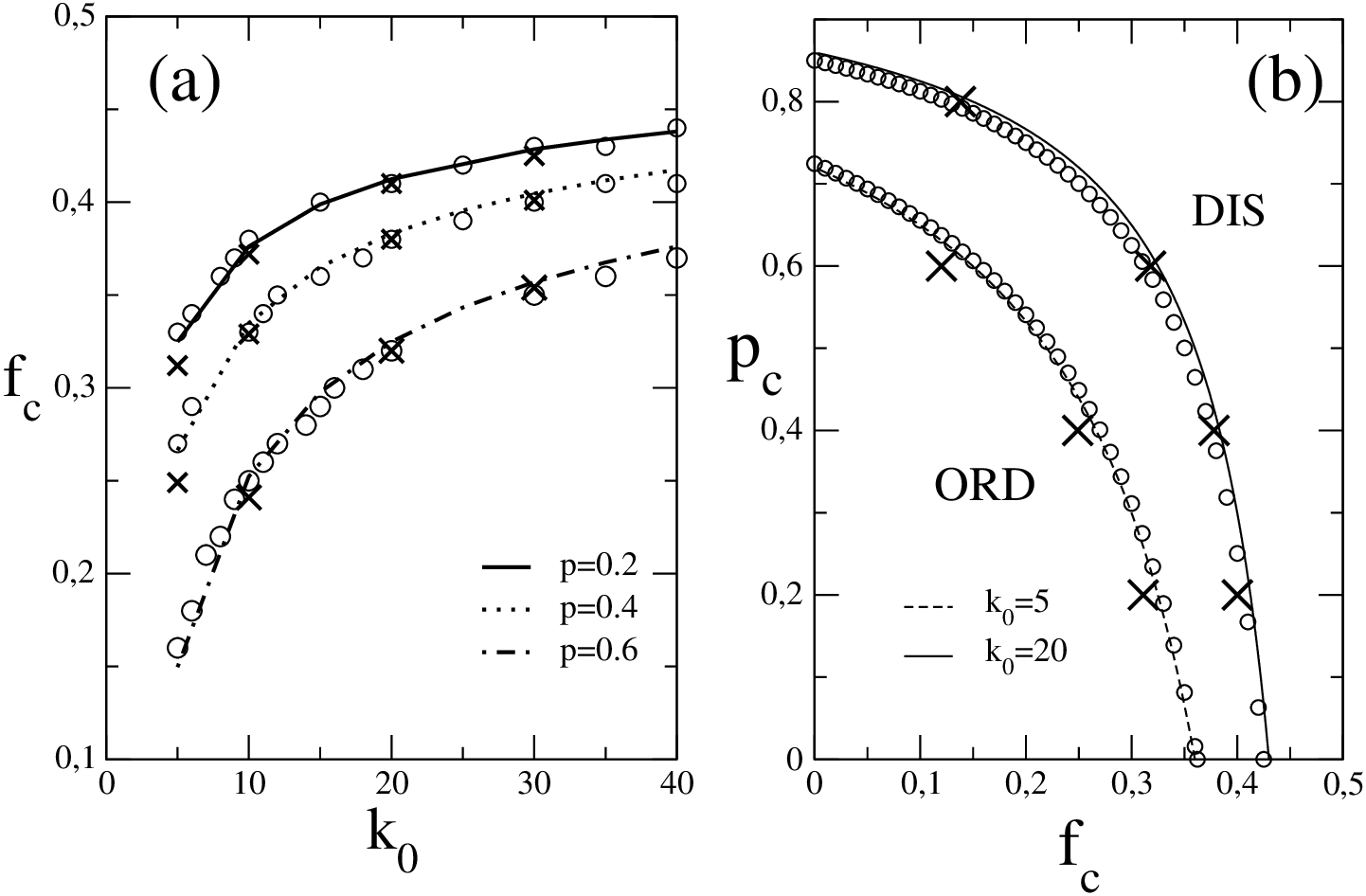,width=9cm,height=5.5cm}
\caption{Panel $(a)$ shows, for the BA case,
  the transition rates $f_c$ versus the minimum degree $k_0$
  for distinct values of $p$. Circles and  lines correspond
  to estimates obtained from Eqs. (\ref{eq4-1}) and (\ref{ucmr}), respectively.
  The
  symbol $\times$ correspond to the numerical values obtained from
  the crossing among fourth-order reduced cumulant for distinct network
  sizes.   Panel  $(b)$  shows the phase diagram $p_c$ versus $f_c$ for distinct
$k_0$s.}
\label{fig4}
\end{figure}

 Now let us examine the phase diagrams.  A comparison  of the results for distinct topologies  is shown
in Figs. \ref{fig2}-\ref{fig4} for distinct values of $p$,  $k_0$ (RR and BA)
and  $\langle k \rangle$ (ER).
 Panels $(a)$ show that all estimates  agree very well for
 high connectivities,  but some discrepancies 
arise for the systems with lower degrees. These trends reveal  not only the reliability
of one-site MFT but
also its accuracy  for the location
of the critical point. 
As in the MFT, the phase transitions are continuous, irrespective of the
lattice topology and the system degree.  This result, contrasts to the observed in the $q-$voter
model, where the inclusion of ``independence'' does turn the phase transition
into a discontinuous one \cite{qvoter}.
 According to \cite{qvoter}, whenever the ``anti-conformism rule''  
is drawn, the chosen site (always) flips its spin provided  all  its $q$ neighbors have opposite spins. 
This interaction rule is slightly different from the usual majority one, in which the spin flip 
depends only on the signal  of the spin neighborhood and 
always there is  a finite 
probability  of the majority rule not to be followed (except to $f=0$).
Also,  increasing the ancillary $p$
does not alter  the discrepancies among these methods.
Finally, we observe that [panels $(b)$] the inclusion of additional noise
shortens the ordered phase and thus the disordered region enlarges.

\section{Conclusions}
 We have investigated the majority vote model in complex networks, in the presence of two distinct kinds of noise.
Our study, through numerical simulations and mean-field theory, considered
both homogeneous and heterogeneous (scale-free) topologies. 
We have derived expressions (through a 
different mean-field approach than Refs. \cite{romualdo,chen1}) 
for the critical point in terms of the network topology and the noise parameter $p$. 
 The resulting expressions work very
well in the regime of high connectivity. The critical behavior
and the set of critical exponents  have been investigated in detail. 
Our numerical results strongly suggest the existence of a common set of critical exponents for the random regular and
Erd\"os-Renyi networks. On the other hand, they are different for heterogeneous
structures,  suggesting a novel universality class for the MV
in scale-free structures. Due to the scarcity of results \cite{ba2},  we believe
that our findings constitute an important step for classifying
the critical behavior in scale-free networks. However, we remark
that further studies are still required to a complete classification. 
In particular, a comparison between
the critical behavior exhibited in other heterogeneous structures, such as the
uncorrelated  configuration model (UCM) \cite{ucm}, can be very interesting.
 Finally, the previous proposal of a discontinuous phase transition 
in the regime
of high connectivity (irrespective of the network topology) 
is discarded \cite{crokidakis}.

\section{Acknowledgement}
We acknowledge the financial support from CNPq,
FAPESP under grant 2018/02405-1.
\newpage
\section*{References}


\end{document}